\documentclass[reprint,aps,pra,longbibliography,noeprint]{revtex4-2}
\usepackage[T1]{fontenc}
\usepackage{graphicx} 
\usepackage[pdftex,linkcolor=blue,pdfborder={0 0 0}]{hyperref} 
\usepackage{calc} 
\usepackage{enumitem} 
\usepackage{parskip}
\usepackage[all]{nowidow} 
\usepackage[protrusion=true,expansion=true]{microtype} 
\usepackage{amsmath}
\usepackage{braket}
\usepackage{bbm} 
\usepackage{amsfonts} 
\usepackage{xcolor}
\usepackage{cleveref} 

\hypersetup{ 	
pdfsubject = {},
pdftitle = {},
pdfauthor = {Jacob Willson}
}

\begin{document} 

\title{Maximum entropy distributions of wavefunctions at thermal equilibrium}

\author{Jacob T. Willson}
\affiliation{Department of Chemistry, Massachusetts Institute of Technology, Cambridge, MA}
\affiliation{Department of Chemistry and Chemical Biology, Harvard University, Cambridge, MA}
\author{Henrik J. Heelweg}
\affiliation{Department of Chemistry, Massachusetts Institute of Technology, Cambridge, MA}
\author{Adam P. Willard}
\affiliation{Department of Chemistry, Massachusetts Institute of Technology, Cambridge, MA}
\email{awillard@mit.edu}

\begin{abstract}
Statistical mechanics reveals that the properties of a macroscopic physical system emerge as an average over an ensemble of statistically independent microscopic subsystems, each occupying a specific microstate. 
In the study of quantum systems, these microstates can be chosen to correspond to the pure state wavefunctions of individual quantum systems.
However, the physical principles that govern the distribution of a pure state wavefunction ensemble, even under conditions of thermal equilibrium, are not well established. 
For instance, the canonical Boltzmann distribution cannot be applied to wavefunctions because they lack a definite energy. 
In this manuscript, we present a maximum entropy principle for the quantum wavefunction ensemble at thermal equilibrium, the so-called Scrooge ensemble. 
We highlight that a constraint on the energy expectation value, or even the shape of the associated eigenstate distribution, fails to yield a valid equilibrium state. 
We find that in addition to these constraints, one must also constrain the measurement entropy to be equal to the R\'enyi divergence of the ensemble with respect to the Gibbs state, indicating that the R\'enyi divergence may have uninvestigated physical importance to thermal equilibrium in quantum systems.
\end{abstract}

\maketitle

\section*{Introduction}

Quantum statistical mechanics is naturally formulated in a basis of energy eigenstates. 
According to this formulation, the observable properties of a macroscopic quantum system represent an average over a weighted ensemble of these eigenstates.  
It therefore stands to reason that before observation, such a system is represented by an equivalently weighted ensemble of wavefunction states that are yet to be collapsed.
Remarkably, the physical principles that determine the weighting of a wavefunction ensemble at thermal equilibrium are not well understood.
Here, we consider the role of entropy maximization in shaping the distribution of wavefunction states within an equilibrium ensemble.

The statistical mechanical formulation of thermal equilibrium is founded on the maximum entropy principle.
This principle states that the equilibrium distribution of system microstates is the one that maximizes entropy subject to the system's physical constraints.\cite{Jaynes1957, Jaynes1957_2}
For quantum systems represented in a basis of eigenstates, when the average energy of the eigenstate ensemble is constrained, entropy maximization yields the well-known Gibbs state, in which eigenstates are Boltzmann distributed, as represented by the density matrix,  
\begin{equation}
    \rho_\mathrm{G} = \frac{e^{-\beta H}}{\mathrm{Tr}\{e^{-\beta H}\}},
    \label{eqn:Gibbs_state}
\end{equation}
where $H$ denotes the system Hamiltonian and $1/\beta=k_\mathrm{B}T$ is the Boltzmann constant times temperature.

A foundational inconsistency arises when this principle is applied to an ensemble of pure state wavefunctions, distributed according to the classical probability distribution $P(\Gamma)$, where $\Gamma=\ket{\psi}\bra{\psi}$ is the density operator representation of the single system wavefunction, $\ket{\psi}$.
Maximizing the entropy of this distribution with the same average energy constraint fails to produce the Gibbs state.\cite{Brody1998, Benvegnu2004, Parfionov2006, Anza2022}
Therefore, if the wavefunction ensemble is governed by a maximum entropy principle, then it must be subject to a different set of physical constraints.
These constraints must shape the distribution in such a way as to be consistent with the Gibbs state upon observation.
In this manuscript, we present the constraints necessary to achieve this consistency and examine why the conventional constraint on average energy, or even on the Gibbs state itself, fails to yield a valid thermal distribution.

It is worthwhile to note that there has been skepticism surrounding the seemingly straightforward extension of statistical mechanical principles to wavefunction ensembles. 
Most notably, it was argued by Jaynes that a distribution of pure state wavefunctions, cannot serve as a valid foundation for statistical mechanics because the state distribution function,  $P(\Gamma)$, does not define the probabilities of mutually exclusive events.\cite{Jaynes1957_2} 
Jaynes argued that wavefunction states are only mutually exclusive if they are orthogonal, and therefore, a quantum state probability distribution is only valid if it is restricted to an orthonormal set of states. 
We argue, however, that although non-orthogonal wavefunction states are observationally indistinguishable, they are formally distinct quantum states.
For example, they are precisely distinguishable (and therefore mutually exclusive) in the numerical simulation of model quantum systems. 
If a statistical mechanical formulation of a simulated ensemble of pure state wavefunctions is possible, then it is, at least in principle, physically meaningful and worthy of study (even if these studies are purely theoretical).

\begin{figure*}[!t]
    \centering
    \includegraphics[width=\textwidth]{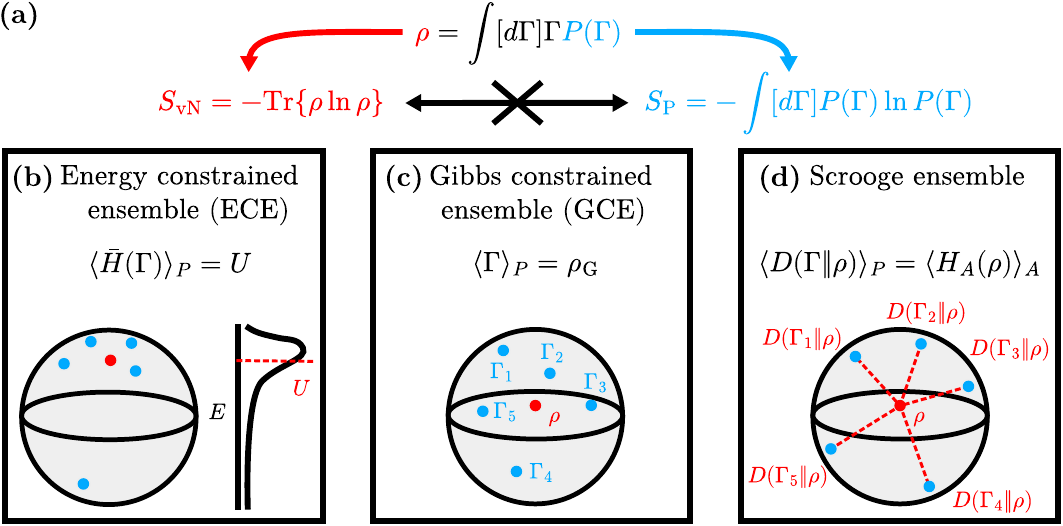}
    \caption{\textbf{Maximum entropy distribution derived from different physical constraints.
    (a)} An ensemble of wavefunctions $P(\Gamma)$ can be compactly represented by a density operator $\rho$. Maximizing the entropy of $P(\Gamma)$ is not equivalent to maximizing the entropy of $\rho$, subject to the same constraint. \textbf{(b)} The energy constrained ensemble (ECE) is the quantum analogue of the classical canonical ensemble, maximizing entropy under an average energy constraint. \textbf{(c)} The Gibbs-state constrained ensemble (GCE) maximizes entropy among distributions that average to the Gibbs state. \textbf{(d)} The Scrooge ensemble maximizes entropy under a constraint on the average R\'enyi divergence $D(\Gamma\|\rho)$ of $\Gamma$ from $\rho$. }
    \label{fig:ensembles}
\end{figure*}

We begin by formalizing the entropy of a quantum state ensemble.
In the standard formulation of quantum statistical mechanics, fundamental thermodynamic properties are defined in relation to the von Neumann entropy,\cite{VonNeumann1927}
\begin{equation}
    S_\mathrm{vN}(\rho) = -\mathrm{Tr}\{\rho \ln \rho\} = -\sum_k p_k \ln p_k,
    \label{eqn:S_vN}
\end{equation}
where $\rho$ is the density operator, and in the second equality the trace is represented by a sum over eigenstates, where $p_k$ denotes the probability of observing a system in the $k$th eigenstate. 
In this study, we define the entropy of a wavefunction ensemble (hereby referred to as the $P$-ensemble entropy) analogously as,
\begin{equation}
    S_P = -\int [d\Gamma] P(\Gamma)\ln P(\Gamma),
    \label{eqn:S_P}
\end{equation}
where the integral is taken over the Haar measure within the state space of all possible pure state wavefunctions, and $P(\Gamma)$ is the probability that the system is in state $\Gamma$.\cite{Dodin2021, Dodin2022}

Although there is no apparent connection between the thermodynamic entropy, $S_\mathrm{vN}$, and the $P$-ensemble entropy, $S_P$, the associated distribution, $P(\Gamma)$ contains useful physical information.
For instance, the ensemble density operator (and therefore all of standard quantum statistical mechanics) is fully encoded by $P(\Gamma)$, as the first moment of the distribution, \textit{i.e.}, 
\begin{equation}
\rho = \int [d\Gamma] \Gamma P(\Gamma).
\label{eqn:firstmoment}
\end{equation}
Furthermore, additional information about the ensemble is contained within the higher moments of $P(\Gamma)$.
This information is, in principle, experimentally accessible, for example, with experimental protocols that utilize quantum information platforms,\cite{Cotler2023, Ippoliti2023, Mark2024, Pilatowsky2024} or single-molecule spectroscopy.\cite{Dodin2019, Dodin2021, Dodin2022}

The mapping from $P(\Gamma)$ to $\rho$ is not one-to-one, \textit{i.e.}, there are many possible wavefunction ensembles capable of representing a given ensemble density matrix, $\rho$.
The first attempt to derive an expression for $P(\Gamma)$ at equilibrium was by Park and Band, who defined an entropy similar to \cref{eqn:S_P}, but defined over all possible density matrices, which they took as the fundamental physical states of quantum systems.\cite{Park1976, Band1976, Park1977, Band1977} 
Their formalism was later refined by Slater, who reduced the probability space to only include pure states;\cite{Slater1991, Slater1992} however, his analysis was restricted to small spin systems.\cite{Slater1992_Minimum, Slater1992_Spin} 
Later, Brody and Hughston considered the full $P$-ensemble, maximizing $S_P$ subject to a constraint on the average energy expectation value. 
In addition, they considered the maximum $S_P$ distribution for a two-level system subject to a constraint to the Gibbs state.\cite{Brody1998, Brody1999, Brody2000, Brody2001, Brody2007} 
Generalizations to $N$-level systems were made soon after for the energy-constrained distribution,\cite{Benvegnu2004, Parfionov2006} and more recently for the Gibbs state-constrained distribution.\cite{Anza2024}\footnote{We catch a minor error in the previously published GCE distribution by providing an alternative proof in section \ref{SI:GCE} of the SI.}

Here, we critically evaluate and build upon these previous efforts with the goal of establishing a maximum entropy principle for the equilibrium wavefunction ensemble.
Before proceeding, we define our criteria for equilibrium. 
Specifically, we will consider $P(\Gamma)$ to be a proper thermal distribution if it satisfies the three criteria originally proposed by Goldstein et al.\cite{Goldstein2006}
These criteria for equilibrium are: 
\begin{enumerate}
\item 
$P(\Gamma)$ yields the Gibbs state, \textit{i.e.}, $\rho_\mathrm{G}=\int [d\Gamma] \Gamma P(\Gamma)$, thus ensuring that $P(\Gamma)$ is consistent with standard quantum statistical mechanics.

\item 
$P(\Gamma)$ is a steady state distribution, \textit{i.e.}, it is invariant under time evolution, which is a simple requirement of equilibrium.

\item 
$P(\Gamma)$ obeys the \textit{hereditary property}. 
That is, if a total system composed of two independent subsystems (\textit{e.g.} a system and a bath) has one subsystem projected out (\textit{e.g.}, by performing a measurement on the bath), then the projected system is time invariant, in the Gibbs state, wnd with wavefunction states distributed according to the same distribution law, $P(\Gamma)$.
The hereditary property ensures that the thermal distribution is inherited by the system when measurements occur in the bath of a system-bath composite. 
\end{enumerate}

It has been previously recognized that these three criteria are mutually satisfied by the so-called Scrooge ensemble, $P_\mathrm{Scr}(\Gamma)$.\cite{Goldstein2006} 
This distribution, as described in detail in the following section, was originally derived based on information-theoretic considerations.
The physical origins of this phenomenological distribution have yet to be established. 
Here, we present a set of physical constraints that yield the Scrooge ensemble as a maximum entropy principle. 

\section*{The Scrooge Ensemble}

The Scrooge ensemble is a probability distribution of pure state wavefunctions, refererenced to a density matrix, $\rho$, defined as, 
\begin{equation}
   P_\mathrm{Scr}(\Gamma) = \frac{1}{(2\pi)^{N-1}}\frac{N!}{\det\rho}\left(\mathrm{Tr}\{\Gamma\rho^{-1}\}\right)^{-(N+1)},
   \label{eqn:Pscr}
\end{equation}
where $N$ is the deminsion of the matrices $\Gamma$ and $\rho$.
This distribution was first derived by Jozsa et al. as the solution to Eq.~\ref{eqn:firstmoment} that minimizes accessible information of the system's wavefunction, thereby being maximally `stingy' with information.\cite{Jozsa1994}
Later, $P_\mathrm{Scr}(\Gamma)$ was independently derived by Goldstein et al., who demonstrated that when $\rho=\rho_\mathrm{G}$, $P_\mathrm{Scr}(\Gamma)$ uniquely satisfies the above three criteria for proper thermal equilibrium.\cite{Goldstein2006} 
Previous work has demonstrated that the Scrooge ensemble has maximum entropy in a certain sense, as it can be obtained from a maximum entropy distribution over unnormalized quantum states (subject to a density matrix constraint), by projecting the unnormalized states to the unit sphere.\cite{Goldstein2006, Mark2024}
However, the physical interpretation of this demonstration is not clear.
If the Scrooge ensemble can be formulated from a maximum entropy principle on normalized wavefunctions, then the associated constraints may guide our understanding of its physical significance.

We now demonstrate that the Scrooge ensemble is the maximum entropy distribution subject to a constraint on the R\'enyi divergence. 
The R\'enyi divergence is a tunable measure of the difference between two probability distributions or density matrices.\cite{Renyi1961, Petz1986} 
Specifically, the R\'enyi divergence between a pure state density matrix $\Gamma$, and an ensemble density matrix, $\rho$, is given by, 
\begin{equation}
    D_\alpha (\Gamma\|\rho) = \frac{1}{\alpha-1}\ln\mathrm{Tr}\{\Gamma\rho^{1-\alpha}\}, \quad \alpha\neq1,
    \label{eqn:Renyi_divergence}
\end{equation}
where $\alpha$ is a tunable parameter capable of shifting the emphasis of the divergence between the features of $\Gamma$ and $\rho$. 
For $\alpha\rightarrow0$, Eq.~\ref{eqn:Renyi_divergence} simply measures the overlap between $\Gamma$ and $\rho$. 
In the limit that $\alpha \rightarrow 1$, the R\'enyi divergence reduces to the Kullback–Leibler divergence, or equivalently to Umegaki's relative entropy\cite{Umegaki1954} ($\lim_{\alpha\rightarrow1}D_\alpha (\Gamma\|\rho) = -\mathrm{Tr}\{\Gamma\ln\rho\}$).
As we increase $\alpha\rightarrow\infty$, the divergence becomes more stringent and sensitive to the largest population difference between $\Gamma$ and $\rho$.\cite{vanErven2014}

We first maximize the $P$-ensemble entropy in Eq.~\ref{eqn:S_P} subject to a generic constraint on the R\'enyi divergence, 
\begin{equation}
    \langle D_\alpha(\Gamma\|\rho)\rangle_P = C(\rho),
    \label{eqn:Dalpha_constraint}
\end{equation}
where $C(\rho)$ is a yet to be specified function of $\rho$. 
We then determine the specific form of $C(\rho)$ that yields the equilibrium Scrooge ensemble.
Using the method of Lagrange multipliers, the maximum entropy distribution is given by,
\begin{equation}
    P_{\alpha\mu}(\Gamma)= \frac{(\mathrm{Tr}\{\Gamma\rho^{1-\alpha}\})^{-\mu}}{\int[d\Gamma](\mathrm{Tr}\{\Gamma\rho^{1-\alpha}\})^{-\mu}}, \quad \alpha \neq 1
    \label{eqn:P_alphamu}
\end{equation}
where $\mu$ is the Lagrange multiplier enforcing the constraint $\langle D_\alpha(\Gamma\|\rho)\rangle_{P_{\alpha\mu}} = C(\rho)$.
A detailed derivation of Eq.~\ref{eqn:P_alphamu} appears in section \ref{SI:Renyi} of the SI.
Comparing Eq.~\ref{eqn:P_alphamu} to Eq.~\ref{eqn:Pscr} reveals that this maximum entropy solution yields the Scrooge ensemble in the case where $\alpha=2$ and $\mu=N+1$. 
Substituting these parameter values into $P_{\alpha \mu}(\Gamma)$, and using the resulting distribution to compute the average in Eq.~\ref{eqn:Dalpha_constraint} provides a solution to the constraint (proof in section \ref{SI:Renyi} of the SI),
\begin{equation}
    C(\rho) = Q(\rho) + H_N - 1,
    \label{eqn:Crho}
\end{equation}
where $Q(\rho)$ is the subentropy of the ensemble density matrix,
\begin{equation}
    Q(\rho) = -\sum_{k=1}^N \frac{p_k^N\ln p_k}{\prod_{i\neq k}(p_k-p_i)},
\end{equation}
and $H_N= \sum_{k=1}^N k^{-1}$ is the $N$-th harmonic number.

To assign physical meaning to this constraint on the R\'enyi divergence, we note that $C(\rho)$ in Eq.~\ref{eqn:Crho} is equivalent to the measurement entropy, averaged over all measurements.
To illustrate this, consider a measurement operator, $A$, with a complete orthonormal basis $\{|a_i\rangle\}$. 
The Shannon entropy of the outcomes of measuring $A$ on $\rho$ is given by,\cite{Shannon1948}
\begin{equation}
    H_A(\rho) = - \sum_{i=1}^N \langle a_i|\rho|a_i\rangle \ln \langle a_i|\rho|a_i\rangle.
\end{equation}
Note that the von Neumann entropy is the minimum value of $H_A(\rho)$, corresponding to a measurement in the eigenbasis of $\rho$. Averaging $H_A(\rho)$ uniformly over all measurement bases (denoted by $\langle \cdot \rangle_A$) yields\cite{Jozsa1994}
\begin{equation}
    \langle H_A(\rho) \rangle_A = Q(\rho) + H_N -1.
\end{equation}
Therefore, the equilibrium distribution of pure state wavefunctions is the maximum entropy distribution that satisfies the constraint, 
\begin{equation}
    \langle D_2(\Gamma\|\rho_\mathrm{G})\rangle_P = \langle H_A(\rho_\mathrm{G}) \rangle_A.
    \label{eqn:thermalconstraint}
\end{equation}

In section \ref{SI:uniqueness} of the SI, we present a proof that the constraint in Eq.~\ref{eqn:thermalconstraint} is the unique constraint on the R\'enyi divergence that is self-consistent, establishing the Scrooge ensemble as an important maximum entropy solution among this class of constraints. 
That is, this particular constraint applied specifically with $\alpha=2$ is the only R\'enyi divergence-based constraint with a maximum entropy principle corresponding to thermal equilibrium.
A consequence of this proof is that it demonstrates that any constrained R\'enyi divergence between the density operator and its composing wavefunctions must yield the Scrooge ensemble in a maximum entropy procedure.
However, it is possible that different physical constraints exist that equivalently yield the Scrooge ensemble as a maximum entropy solution.

The constraint in Eq.~\ref{eqn:thermalconstraint} includes a hierarchy of indirect constraints.
This includes a constraint that the first moment of $P(\Gamma)$ is equal to the Gibbs state, $\rho_\mathrm{G}$, which, in turn, includes an indirect constraint on the average energy of the wavefunction ensemble.
In the following sections, we ascend this hierarchy of constraints and evaluate the consequences for the associated maximum entropy principles.

\section*{Energy-constrained ensemble}

\begin{figure}
    \centering
    \includegraphics[width=\columnwidth]{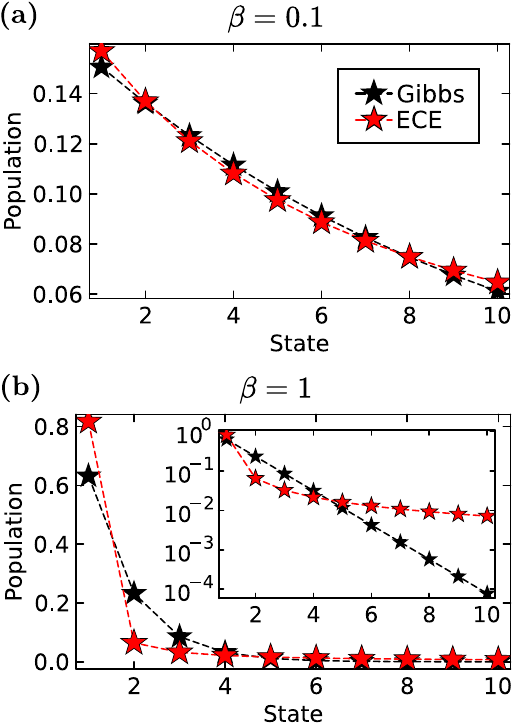}
    \caption{\textbf{The populations of the ECE (red stars) do not agree with the Gibbs state (black stars).} \textbf{(a)} The thermal populations at high temperature are similar. \textbf{(b)} At low temperature, the ground state and higher excited state populations are increased in the ECE. Inset shows populations on a logarithmic scale. Results were calculated for the Hamiltonian spectrum $\sigma\{H\}=\{k\}_{k=0}^{9}$.}
    \label{fig:thermal_pops}
\end{figure}

We now consider the maximum entropy distribution subject to a constraint on the average energy expectation value.
We will refer to this maximum entropy distribution as the energy-constrained ensemble (ECE).
For a general Hamiltonian $H$ with finite dimension $N$, the ECE is obtained by maximizing $S_P$ subject to an average energy constraint, yielding a Boltzmann-like distribution over wavefunctions (proof in section \ref{SI:ECE} of the SI), 
\begin{equation}
    P_\mathrm{ECE}(\Gamma) = \frac{\exp(-\beta'\bar{H}(\Gamma))}{\int[d\Gamma] \exp(-\beta'\bar{H}(\Gamma))}
    \label{eqn:P_Gamma}
\end{equation}
where $\bar{H}(\Gamma) = \mathrm{Tr}\{H\Gamma\}$ is an energy associated with the state $\Gamma$, and $\beta'$ is a statistical inverse temperature that depends on the average energy. 
Notably, $\beta'$ differs from the thermodynamic inverse temperature, $\beta$.
However, the two quantities can be related through the energy constraint expression (proof in section \ref{SI:ECE} of the SI),\cite{Harish1957, Itzykson1980, McSwiggen2021}
\begin{equation}
    \frac{\sum_i E_i \alpha_i}{\sum_i \alpha_i} + \frac{N-1}{\beta'} = \frac{\sum_i E_i e^{-\beta E_i}}{\sum_i e^{-\beta E_i}},
    \label{eqn:betaprime_constraint_simplified}
\end{equation}
where 
\begin{equation}
    \alpha_i = \frac{e^{-\beta' E_i}}{\prod_{j \neq i}(\beta'E_j -\beta'E_i)},
\end{equation}
and $E_i$ denotes the eigenvalue of the $i$th energy eigenstate.

The density operator of the ECE, $\rho_\mathrm{ECE}=\int [d\Gamma]\Gamma P_\mathrm{ECE}(\Gamma)$, does not coincide with the Gibbs state, demonstrating that the average energy constraint on the space of all wavefunctions is distinct from the same constraint on eigenstates.\cite{Brody1998, Benvegnu2004, Parfionov2006, Anza2022} The population of the $k$-th state in the ECE is given by (proof in  section \ref{SI:ECE_operator} of the SI)
\begin{equation}
    p_{k}^\mathrm{ECE}:=\langle \Gamma_{kk} \rangle_{P_\mathrm{ECE}} = \frac{\alpha_k + \sum_{i\neq k} \frac{\alpha_k  + \alpha_i}{\beta'E_k - \beta'E_i}}{\sum_{i=1}^N \alpha_i},
    \label{eqn:P_thermal_state}
\end{equation}
where $\langle \cdot \rangle_P$ denotes the average over $P(\Gamma)$. \Cref{fig:thermal_pops} shows that whilst the thermal populations of the ECE and Gibbs state are similar at high temperatures, they become significantly different at low temperatures. Furthermore, the differences increase with system size, where the ECE exhibits a condensation into the ground state in the thermodynamic limit (shown in  section \ref{SI:condensation} of the SI).\cite{Fine2009, Hantschel2011}

Since the ECE does not coincide with the ubiquitous Gibbs state, it is not a valid thermal distribution of wavefunctions. 
As noted elsewhere,\cite{Alonso2015} the ECE partition function is not factorizable and therefore fails to produce extensive thermodynamic properties.

\section*{Gibbs-constrained ensemble}

We now consider the consequences of maximizing the entropy subject to a constraint to the Gibbs state.
We refer to this distribution as the Gibbs-constrained ensemble (GCE).
As we show in  section \ref{SI:GCE} of the SI, the GCE is given by,
\begin{equation}
    P_\mathrm{GCE}(\Gamma) = \frac{ \exp(-\sum_i \mu_i \Gamma_{ii})}{\int [d\Gamma]\ \exp(-\sum_i \mu_i \Gamma_{ii})} ,
    \label{eqn:P_GCE}
\end{equation}
where $\{\mu_i\}_{i=1}^N$ is a set of parameters that satisfy a set of $N$ coupled equations,
\begin{equation}
    p_k = \frac{\tilde\alpha_k - \sum_{i\neq k} \frac{\tilde\alpha_i + \tilde\alpha_k}{\mu_i - \mu_k}}{\sum_i \tilde\alpha_i}, \quad \tilde\alpha_i = \frac{e^{-\mu_i}}{\prod_{j \neq i} (\mu_j - \mu_i)},
    \label{eqn:GCE_LM}
\end{equation}
where $p_k = (\rho_\mathrm{G})_{kk}$ are the populations of the Gibbs state. At high temperatures, the GCE approaches the ECE and becomes Boltzmann-like with $\mu_i \approx \beta'E_i$. At low temperatures, the parameters $\{\mu_i\}$ deviate from the ECE to maintain the Gibbs state distribution and thereby avoid the unphysical ground state condensation.

Although the Gibbs state is preserved in the GCE, we demonstrate that the GCE violates the hereditary property and thus cannot be a valid thermal distribution. The hereditary property requires the projected ensemble on the system (\textit{e.g.}, in a system-bath composite) to be the same as the thermal distribution on the system.\cite{Goldstein2006} 
More concretely, let $|\Psi\rangle\in\mathcal{H}_1\otimes\mathcal{H}_2$ be a wavefunction sampled from the thermal distribution $P(\Gamma)$ on the composite Hilbert space of $\mathcal{H}_1$ and $\mathcal{H}_2$. Then the projected ensemble on $\mathcal{H}_1$ given a measurement basis $\{|q_2\rangle\}$ for $\mathcal{H}_2$ is
\begin{equation}
|\psi_1\rangle =\frac{\langle Q_2|\Psi\rangle}{\|\langle Q_2|\Psi\rangle\|} \in \mathcal{H}_1,
\end{equation}
where $Q_2$ is a random measurement basis state with distribution according to the Born rule,
\begin{equation}
   \mathbb{P}(Q_2=q_2) = \|\langle q_2|\Psi\rangle\|^2.
\end{equation}
The hereditary property requires that the projected ensemble distribution $P_\mathrm{proj}(|\psi_1\rangle\langle\psi_1|)$ be equal to the thermal distribution $P(|\psi_1\rangle\langle\psi_1|)$ on $\mathcal{H}_1$.

\begin{figure}
    \centering
    \includegraphics[width=\columnwidth]{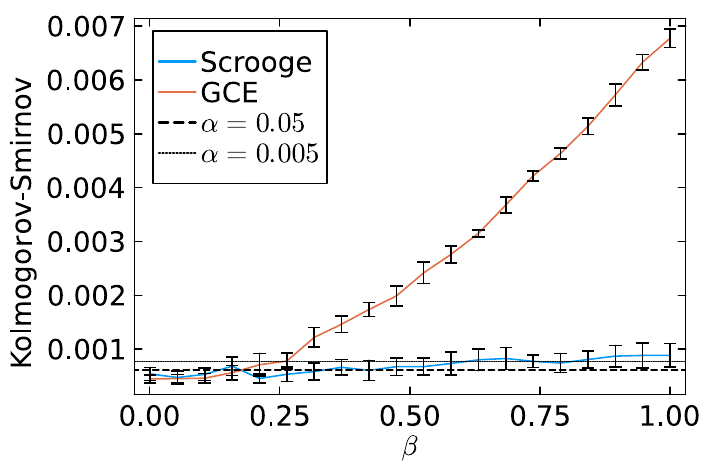}
    \caption{\textbf{The hereditary property is violated by the GCE and satisfied by the Scrooge ensemble.} The Kolmogorov-Smirnov (KS) statistic of the projected ensemble and thermal distribution of the system as a function of inverse temperature $\beta$. The GCE satisfies the KS test at high temperatures for the significance levels of $\alpha =  5\%$ and $0.5\%$, but fails at low temperatures. The Scrooge ensemble satisfies the hereditary property for all temperatures.\cite{Goldstein2006} Results are calculated by Monte Carlo sampling, with the system and bath spectra given by $\sigma\{H\} = \{0,1\}$. Error bars are standard error in the mean over $10$ trials with $10^7$ samples.}
    \label{fig:KS_test}
\end{figure}

We characterise the difference between these two distributions by the Kolmogorov-Smirnov (KS) statistic\cite{Knuth1997} of the density of states (DOS),
\begin{equation}
    D = \sup_E |P_\mathrm{GCE}^\mathrm{proj}(E)-P_\mathrm{GCE}(E)|,
\end{equation}
where $P_\mathrm{GCE}(E)=\int[d\Gamma]P_\mathrm{GCE}(\Gamma)\delta(\bar{H}(\Gamma)-E)$ is the DOS of the GCE on the system, and $P_\mathrm{GCE}^\mathrm{proj}(E)$ is the projected ensemble DOS. If the KS statistic exceeds the threshold $D(p)=\sqrt{-\ln(p/2)/N_\mathrm{samp}}$, where $N_\mathrm{samp}$ is the number of sampled wavefunctions, then the null hypothesis that both distributions are the same is rejected at significance level~$\alpha$. \Cref{fig:KS_test} shows the KS statistic as a function of $\beta$, with the corresponding thresholds for $\alpha =  5\%$ and $0.5\%$. At high temperatures, the distributions are similar due to both converging to the uniform Haar ensemble. At lower temperatures, the KS statistic significantly exceeds the given thresholds and the hereditary property is violated in the GCE. These numerical results support earlier conjectures that there is no reason to expect that the GCE exhibits the hereditary property.\cite{Goldstein2006}

\section*{Discussion}

We have shown that relaxing the R\'enyi divergence constraint to more traditional constraints on the average energy or Gibbs state does not yield valid thermal distributions of wavefunctions. This finding provides the R\'enyi divergence with newfound importance in the statistical mechanics of wavefunction ensembles.

The R\'enyi divergence $D_2(\Gamma\|\rho)$ has interesting properties beyond being the necessary maximum entropy constraint for the Scrooge ensemble.
First, the divergence with $\alpha=2$ is the maximal value that satisfies the Data Processing Inequality (DPI): Given two density operators $\sigma, \rho$ and a completely positive trace-preserving map $\Lambda$, then $D_\alpha(\Lambda(\sigma)\|\Lambda(\rho))\leq D_\alpha(\sigma\|\rho)$.\cite{Ando1979, Audenaert2015} 
The DPI must be satisfied by physically useful divergences, since applying quantum operations cannot make it easier to distinguish the initial states.
Since larger values of $\alpha$ apply more stringent comparisons between the distributions,\cite{vanErven2014} $D_2(\Gamma\|\rho)$ is the most stringent divergence that is physically valid. Second, the parameter $\alpha$ characterizes the averaging function in the R\'enyi entropy $H_\alpha$ and R\'enyi divergence i.e., $H_\alpha = \langle -\ln p\rangle_{g_\alpha} = g_\alpha^{-1}(\sum_k p_k g_\alpha(-\ln p_k))$, where $g_\alpha(x) = e^{(1-\alpha)x}$.\cite{Renyi1961} The case $\alpha=2$ corresponds to a simple exponential decay weighting $g_2(x)=e^{-x}$. Third, the divergence $D_\alpha(\Gamma\|\rho)$ is interpreted as the expected increase in surprisal when using the approximate distribution $\rho$, if the true distribution is $\Gamma$. This set-up mirrors an experiment where measurements are performed on an unknown $\Gamma$, drawn from an ensemble $P(\Gamma)$ with known $\rho$. Therefore, the constraint $\langle D_\alpha(\Gamma\|\rho)\rangle_P$ captures the excess surprisal of the underlying distribution $P(\Gamma)$, when we only know $\rho$.

The Scrooge ensemble has been previously invoked as the maximum entropy distribution subject to a density operator constraint (i.e., the GCE).\cite{Mark2024, Mandal2026, Sherry2026} This claim is only true in the limit of infinite temperature, where the Scrooge ensemble and GCE both approach the uniform random Haar ensemble of wavefunctions. In general, the Scrooge ensemble has less entropy than the GCE for $\beta>0$, and in the thermodynamic limit ($N \rightarrow \infty$) (proof in  section \ref{SI:entropy} of the SI). To remedy this discrepancy and ensure that the Scrooge ensemble has maximum entropy subject to a Gibbs state constraint, some approaches have redefined the entropy of an ensemble as the (negative) accessible information, instead of the Shannon entropy of the ensemble.\cite{Liu2024, Chang2025, Mok2026}

The maximum entropy constraint of the Scrooge ensemble currently lacks a strong a priori physical justification. Due to the uniqueness of the constraint $\langle D_2(\Gamma\|\rho)\rangle_P= \langle H_A(\rho) \rangle_A$ amongst the class of divergence constraints, justifying it reduces to rationalizing why the average excess surprisal, or loosely speaking, the `distance' of wavefunctions from their average $\rho$, should depend on $\rho$ in an ensemble of non-interacting and statistically independent systems at equilibrium.

It remains an open theoretical challenge to justify the maximization of the von Neumann entropy before the ensemble entropy. This entropy hierarchy is assumed in both the GCE and Scrooge ensemble by constraining the distributions to average to the Gibbs state. Ideally, the Gibbs state should emerge out of the constraints on the wavefunction distribution, instead of being inserted as a constraint itself. 

\section*{Conclusion}

In conclusion, we have proven that the wavefunction ensemble at thermal equilibrium, the Scrooge ensemble, satisfies the maximum entropy principle under the constraint that the average R\'enyi divergence of the wavefunctions is equal to the average measurement entropy over all measurements.
This unusual constraint is necessary since more traditional constraints on the average energy expectation value or constraining populations to match the Gibbs state do not yield valid thermal distributions of wavefunctions, even in the thermodynamic limit. 
The appearance of the R\'enyi divergence in the maximum entropy foundations of the Scrooge ensemble suggests that it may have uninvestigated physical importance to thermal equilibrium in quantum systems.

\bibliography{bib}

\setcounter{section}{0}
\setcounter{figure}{0}
\setcounter{table}{0}
\setcounter{equation}{0}

\renewcommand{\thefigure}{S\arabic{figure}}
\renewcommand{\thetable}{S\arabic{table}}
\renewcommand{\theequation}{S\arabic{equation}}
\renewcommand{\thesection}{S\arabic{section}}

\section{Ensemble entropy maximization}
\label{SI:entmax}

In this section, we derive maximum entropy distributions under three different sets of constraints. The first case is an average energy expectation value constraint, which mirrors the classical derivation of the Boltzmann distribution and yields a Boltzmann-like distribution. In this case, the von Neumann entropy is not maximized, and the resultant density operator does not agree with the Gibbs state. The second case is a Gibbs-state constraint on the populations of the density matrix, which ensures that the von Neumann entropy is maximized. The third case is a R\'enyi divergence constraint, which generates the Scrooge ensemble for $\alpha=2$ and constraint $C(\rho) = \langle H_A(\rho)\rangle_A$.

The information-theoretic approach to statistical mechanics asserts that the most unbiased representation of our knowledge of the state of a system is obtained by maximizing the Shannon entropy subject to the known constraints.\cite{Jaynes1957, Jaynes1957_2} The Shannon entropy over a discrete set of wavefunctions $\mathcal{G}$ is given by
\begin{equation}
    S_P=-\sum_{\Gamma_j\in\mathcal{G}} p_j \ln p_j,
    \label{Seqn:S_P}
\end{equation}
where $p_j$ is the probability of state $\Gamma_j$. The constraints on an ensemble of wavefunctions are generally formulated through expectation values,
\begin{equation}
    \langle f_k \rangle = \sum_{\Gamma_j\in\mathcal{G}} p_j f_k(\Gamma_j),
    \label{Seqn:constraints}
\end{equation}
where $\{f_k\}_{k=1}^M$ is a set of observables with known constraints. The entropy \cref{Seqn:S_P} can be maximized under the constraints \cref{Seqn:constraints} through the method of Lagrange multipliers to yield the maximum entropy probability distribution,
\begin{equation}
    p_j = \frac{\exp \left(-\sum_{k=1}^M\mu_kf_k(\Gamma_j)\right)}{\sum_{\Gamma_j\in\mathcal{G}} \exp\left(-\sum_{k=1}^M\mu_kf_k(\Gamma_j)\right)},
\end{equation}
where $\{\mu_k\}_{k=1}^M$ is a set of Lagrange multipliers that are chosen to satisfy the constraints. In the limit that the discrete set of wavefunctions becomes dense and uniformly distributed, we obtain the continuous probability density,
\begin{equation}
    P(\Gamma) = \frac{\exp \left(-\sum_{k=1}^M\mu_kf_k(\Gamma)\right)}{\int[d\Gamma] \exp\left(-\sum_{k=1}^M\mu_kf_k(\Gamma)\right)},
    \label{Seqn:P_constraints}
\end{equation}
where $\int [d\Gamma]$ integrates over the Haar measure for wavefunctions.

\subsection{Energy constraint}
\label{SI:ECE}

The average energy constraint defines the analogue of the canonical ensemble over wavefunction state space. The constraint and corresponding observable are given by 
\begin{gather}
    \langle \bar{H}(\Gamma)\rangle =U, \\
    f(\Gamma) \rightarrow \bar{H}(\Gamma)=\mathrm{Tr}\{H\Gamma\}, \label{Seqn:energy_constraint}
\end{gather}
where $H$ is the Hamiltonian, and $U$ is the average energy. Substituting \cref{Seqn:energy_constraint} into \cref{Seqn:P_constraints} yields the energy constrained ensemble (ECE) distribution,
\begin{equation}
    P_\mathrm{ECE}(\Gamma) = \frac{\exp \left(-\beta' \bar{H}(\Gamma)\right)}{\int[d\Gamma] \exp\left(-\beta' \bar{H}(\Gamma)\right)},
\end{equation}
where $\beta'$ is a Lagrange multiplier that can be interpreted as the statistical inverse temperature over all wavefunctions. The value of $\beta'$ is set through the energy constraint, which can be rewritten in terms of the partition function over state space $\mathcal{Z}_\mathrm{ECE}$,
\begin{equation}
    U = -\frac{\partial}{\partial\beta'}\ln\left(\mathcal{Z}_\mathrm{ECE}\right), \quad \mathcal{Z}_\mathrm{ECE} = \int[d\Gamma]\exp\left(-\beta'\bar{H}(\Gamma)\right).
    \label{Seqn:U}
\end{equation}
The partition function can be evaluated by treating each wavefunction $\Gamma$ as an element of the unitary orbit $\{\Gamma=U\Gamma_0U^\dagger|\ U \in U(N)\}$ of an initial wavefunction $\Gamma_0$. The integral over all wavefunctions can then be replaced by an integral over the unitary group $U(N)$, and the resulting equation can be evaluated using the HCIZ integral,\cite{Harish1957, Itzykson1980, McSwiggen2021}
\begin{align}
    \frac{\int[d\Gamma]e^{-\beta'\mathrm{Tr}\{H\Gamma\}}}{\int[d\Gamma]}&=\int_{U(N)}[dU] \exp\left(-\beta'\mathrm{Tr}\{HU\Gamma_0U^\dagger\}\right),  \\
    &=\left(\prod_{i=1}^{N-1}i!\right) \frac{\mathrm{det}[e^{-\beta'E_i\lambda_j}]_{1\leq i,j\leq N}}{(-\beta')^{\frac{N^2-N}{2}}\Delta(H)\Delta(\Gamma_0)},
\end{align}
where $\int[d\Gamma]=\frac{(2\pi)^{N-1}}{(N-1)!}$ is the state space volume, $\{E_i\}_{i=1}^N$ are the eigenvalues of $H$, $\{\lambda_i\}_{i=1}^N$ are the eigenvalues of $\Gamma_0$, and $\Delta(H)$ is the Vandermonde determinant of $H$,
\begin{equation}
    \Delta(H)=
    \begin{vmatrix}
        1 & E_1 & E_1^2 & \cdots  & E_1^{N-1} \\

        \vdots & \vdots & \vdots & \ddots & \vdots \\

        1 & E_N & E_N^2 & \cdots  & E_N^{N-1} \\
    \end{vmatrix},
\end{equation}
and similarly for $\Delta(\Gamma_0)$. Assuming we have a non-degenerate set of energies $\{E_i\}_{i=1}^N$, and setting the initial wavefunction to $\Gamma_0 = \mathrm{diag}(1, 0, \dots, 0)$, we can apply L'H\^opital's rule for the vanishing $\{\lambda_i\}_{i=2}^N$ to yield 
\begin{equation}
    \mathcal{Z}_\mathrm{ECE} = \left(\frac{-2\pi}{\beta'}\right)^{N-1}\frac{\Delta'(H;\beta')}{\Delta(H)},
    \label{Seqn:partition_detratio}
\end{equation}
where 
\begin{equation}
    \Delta'(H;\beta') = \begin{vmatrix}
        1 & E_1 & E_1^2 & \cdots & E_1^{N-2} & e^{-\beta'E_1} \\

        \vdots & \vdots & \vdots & \ddots & \vdots & \vdots \\

        1 & E_N & E_N^2 & \cdots & E_N^{N-2} & e^{-\beta'E_N} \\
    \end{vmatrix}.
\end{equation}
For a degenerate set of energies, further applications of L'H\^opital's rule can be performed without difficulty. The ratio of determinants in \cref{Seqn:partition_detratio} can be simplified to yield an analytic form of the partition function,
\begin{equation}
     \mathcal{Z}_\mathrm{ECE} = (2\pi)^{N-1}\sum_{i=1}^N \alpha_i, \quad \alpha_i = \frac{e^{-\beta'E_i}}{\prod_{j\neq i}(\beta'E_j-\beta'E_i)}.
     \label{Seqn:Z_analytic}
\end{equation}
Substituting \cref{Seqn:Z_analytic} into the energy constraint \cref{Seqn:U} yields
\begin{equation}
    U = -\frac{\partial}{\partial\beta'}\ln\left(\sum_{i=1}^N\alpha_i\right)=\frac{\sum_{i=1}^N E_i \alpha_i}{\sum_{i=1}^N \alpha_i} + \frac{N-1}{\beta'},
\end{equation}
which implicitly defines $\beta'$, given $H$ and $U$. Equating the energy $U$ to the thermal energy of the Gibbs state (thereby applying the same constraint on the ensemble entropy and von Neumann entropy) gives
\begin{equation}
    \frac{\sum_i E_i \alpha_i}{\sum_i \alpha_i} + \frac{N-1}{\beta'} = \frac{\sum_i E_i e^{-\beta E_i}}{\sum_i e^{-\beta E_i}},
    \label{Seqn:betaprime_constraint_simplified}
\end{equation}
which allows conversion between $\beta'$ and the thermodynamic temperature $\beta$, as presented in the main text.

\begin{figure}
    \centering
    \includegraphics[width=\columnwidth]{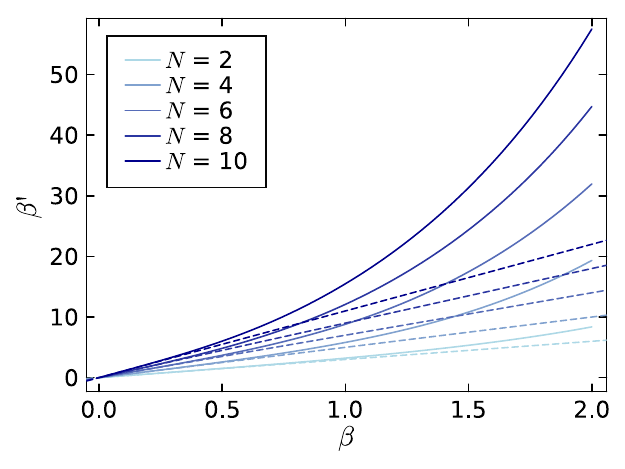}
    \caption{\textbf{The statistical temperature $\beta'$ of the ECE is not the thermodynamic temperature $\beta$.} Results are calculated for Hamiltonians with equally spaced eigenvalues $\sigma\{H\}=\{k\}_{k=0}^{N-1}$, where $N$ is the dimension. Dashed lines are the high temperature limit $\beta'=(N+1)\beta$.}
    \label{fig:betaprime}
\end{figure}

\Cref{fig:betaprime} shows $\beta'(\beta)$ for a set of Hamiltonians with equally spaced eigenvalues $\sigma\{H\}=\{k\}_{k=0}^{N-1}$. At high temperatures, the relationship is approximately linear with $\beta'=(N+1)\beta$, for any Hamiltonian.

\subsection{Gibbs-state constraint}
\label{SI:GCE}

Applying the Gibbs-state constraint in \cref{Seqn:P_constraints} gives the maximum entropy distribution that also maximizes the von Neumann entropy. The Gibbs-state constraint consists of $N$ population constraints and $N(N-1)$ real-valued coherence constraints,
\begin{gather}
     \langle \Gamma_{nn} \rangle=p_n =\frac{e^{-\beta E_n}}{\sum_{i=1}^N e^{-\beta E_i}}, \quad n \in \{1,2,\cdots, N\},\\  \langle \Gamma_{mn}^\mathrm{R}\rangle=\langle\Gamma_{mn}^\mathrm{I}\rangle=0, \quad m > n,
\end{gather}
where $\langle \Gamma_{nn} \rangle$ are the populations, and $\langle \Gamma_{mn}^\mathrm{R}\rangle$ and $\langle \Gamma_{mn}^\mathrm{I}\rangle$ are the real and imaginary parts of the coherences. The corresponding set of $N^2$ observables is given by
\begin{equation}
    \{f_k(\Gamma)\}_k \rightarrow 
    \begin{cases}
        \Gamma_{nn}= \langle n|\Gamma| n\rangle, \\
        \Gamma_{mn}^\mathrm{R}=\mathrm{Re}\{\langle m|\Gamma| n\rangle\}, \\
        \Gamma_{mn}^\mathrm{I}=\mathrm{Im}\{\langle m|\Gamma| n\rangle\},
    \end{cases}
\end{equation}
which can be substituted into \cref{Seqn:P_constraints} to yield
\begin{equation}
    P_\mathrm{GCE}(\Gamma) = \frac{e^{-\sum_{n=1}^N\mu_n \Gamma_{nn}-\sum_{n<m}\eta^\mathrm{R}_{mn}\Gamma_{mn}^\mathrm{R}-\sum_{n<m}\eta^\mathrm{I}_{mn}\Gamma_{mn}^\mathrm{I}}}{\int[d\Gamma] e^{-\sum_{n=1}^N\mu_n \Gamma_{nn}-\sum_{n<m}\eta^\mathrm{R}_{mn}\Gamma_{mn}^\mathrm{R}-\sum_{n<m}\eta^\mathrm{I}_{mn}\Gamma_{mn}^\mathrm{I}}},
    \label{Seqn:P_Gibbs_allLMs}
\end{equation}
where $\{\mu_n\}_{n=1}^N$, $\{\eta_{mn}^\mathrm{R}\}_{m>n}$ and $\{\eta_{mn}^\mathrm{I}\}_{m>n}$ are the Lagrange multipliers for the population and coherence constraints, respectively.

We will now show that the coherence Lagrange multipliers vanish due to the structure of state space. For every state $\Gamma$, we define the real reflection $\Gamma^{(mn,\mathrm{R})}$ by inverting the sign of the real part of the coherence $\Gamma_{mn}^\mathrm{R} \rightarrow -\Gamma_{mn}^\mathrm{R}$. The relative probability of the state $\Gamma$ to its real reflected state is given by
\begin{equation}
    \frac{P_\mathrm{GCE}(\Gamma)}{P_\mathrm{GCE}(\Gamma^{(mn,\mathrm{R})})}= \exp\left(-2\eta^\mathrm{R}_{mn}\Gamma_{mn}^\mathrm{R}\right).
    \label{Seqn:relative_prob}
\end{equation}
The real coherence constraint can be written over the states $\Gamma$ or over the real reflected states $\Gamma^{(mn,\mathrm{R})}$ as
\begin{equation}
    \langle \Gamma_{mn}^\mathrm{R}\rangle = \int [d\Gamma]P(\Gamma)\Gamma_{mn}^\mathrm{R} = \int [d\Gamma]P(\Gamma^{(mn, \mathrm{R})})(-\Gamma_{mn}^\mathrm{R}).
    \label{Seqn:reflected_constraint}
\end{equation}
Adding both expressions of \cref{Seqn:reflected_constraint}, and applying \cref{Seqn:relative_prob} yields
\begin{align}
    \langle \Gamma_{mn}^\mathrm{R}\rangle &= \frac{1}{2}\int [d\Gamma]\Gamma_{mn}^\mathrm{R}\left(P(\Gamma)-P(\Gamma^{(mn, \mathrm{R})})\right) \\
    &= \frac{1}{2}\int [d\Gamma]\Gamma_{mn}^\mathrm{R} P(\Gamma)\left(1-\exp(2\eta^\mathrm{R}_{mn}\Gamma_{mn}^\mathrm{R})\right),
\end{align}
which can only satisfy $\langle \Gamma_{mn}^\mathrm{R}\rangle=0$ if 
\begin{equation}
    \eta^\mathrm{R}_{mn}=0.
\end{equation}
Therefore, all $\{\eta_{mn}^\mathrm{R}\}_{m>n}$ must vanish. By an identical proof, all $\{\eta_{mn}^\mathrm{I}\}_{m>n}$ can be shown to vanish by considering the imaginary reflection $\Gamma^{(mn,\mathrm{I})}$, defined by $\Gamma_{mn}^\mathrm{I} \rightarrow -\Gamma_{mn}^\mathrm{I}$. Consequently, the Gibbs-state constrained ensemble (GCE) distribution \cref{Seqn:P_Gibbs_allLMs} reduces to
\begin{equation}
    P_\mathrm{GCE}(\Gamma) = \frac{\exp \left(-\sum_{n=1}^N\mu_n \Gamma_{nn}\right)}{\int[d\Gamma] \exp \left(-\sum_{n=1}^N\mu_n \Gamma_{nn}\right)}.
\end{equation}

\begin{figure}
    \centering
    \includegraphics[width=\columnwidth]{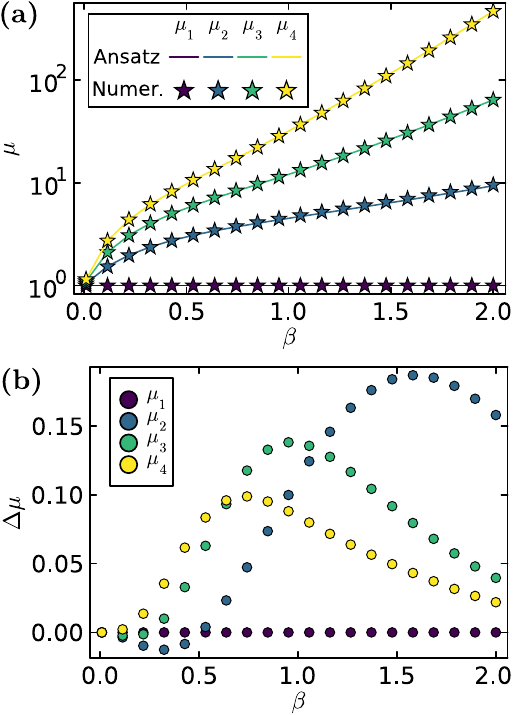}
    \caption{\textbf{The Lagrange multipliers $\mu_i$ of the GCE can be well-approximated by an ansatz. (a)} The numerical solution to \cref{eqn:mu_equations} (stars) is well approximated by the ansatz \cref{eqn:mu_ansatz} (lines). \textbf{(b)} The difference between the numerical solution and ansatz vanishes at high and low temperatures. At intermediate temperatures, the difference is small compared to the absolute value of $\mu_i$. The results are calculated for a Hamiltonian spectrum $\sigma\{H\} = \{0,1,2,3\}$.}
    \label{fig:mus}
\end{figure}

Lastly, we must obtain the values of the population Lagrange multipliers $\{\mu_n\}_{n=1}^N$ from the population constraints, which can be rewritten as 
\begin{equation}
    \langle \Gamma_{nn}\rangle=-\frac{\partial}{\partial\mu_n}\ln \mathcal{Z}_\mathrm{GCE}, \quad \mathcal{Z}_\mathrm{GCE} = \int[d\Gamma] \exp(-\mathrm{Tr}\{M\Gamma\})
    \label{Seqn:GC_pop_constraint}
\end{equation}
where we have defined $M = \mathrm{diag}(\mu_1, \mu_2, \dots, \mu_N)$. The integral over state space in \cref{Seqn:GC_pop_constraint} can be converted into the HCIZ integral over the unitary group (as described in the previous section),
\begin{align}
    \frac{\int[d\Gamma] \exp(-\mathrm{Tr}\{M\Gamma\})}{\int[d\Gamma]}&=\int_{U(N)}[dU] \exp\left(-\mathrm{Tr}\{MU\Gamma_0U^\dagger\}\right), \\
    &=\left(\prod_{i=1}^{N-1}i!\right) \frac{\mathrm{det}[e^{-\mu_i\lambda_j}]_{1\leq i,j\leq N}}{(-1)^{\frac{N^2-N}{2}}\Delta(M)\Delta(\Gamma_0)},
\end{align}
where $\{\lambda_i\}_{i=1}^N$ are the eigenvalues of the initial wavefunction state $\Gamma_0$, and $\Delta(M)$ is the Vandermonde determinant of $M$,
\begin{equation}
    \Delta(M)=
    \begin{vmatrix}
        1 & \mu_1 & \mu_1^2 & \cdots  & \mu_1^{N-1} \\

        \vdots & \vdots & \vdots & \ddots & \vdots \\

        1 & \mu_N & \mu_N^2 & \cdots  & \mu_N^{N-1} \\
    \end{vmatrix},
\end{equation}
and similarly for $\Delta(\Gamma_0)$. Assuming that $\{\mu_n\}_{n=1}^N$ is non-degenerate (which is true if $\{E_i\}_{i=1}^N$ is non-degenerate), and setting the initial wavefunction to $\Gamma_0 = \mathrm{diag}(1, 0, \dots, 0)$, we can apply L'H\^opital's rule for the vanishing $\{\lambda_i\}_{i=2}^N$ to yield 
\begin{equation}
    \mathcal{Z}_\mathrm{GCE}=(-2\pi)^{N-1}\frac{\tilde\Delta(M)}{\Delta(M)},
    \label{Seqn:GC_partition_detratio}
\end{equation}
where
\begin{equation}
    \tilde\Delta(M)=
    \begin{vmatrix}
        1 & \mu_1 & \mu_1^2 & \cdots & \mu_1^{N-2} & e^{-\mu_1} \\

        \vdots & \vdots & \vdots & \ddots & \vdots & \vdots \\

        1 & \mu_N & \mu_N^2 & \cdots & \mu_N^{N-2} & e^{-\mu_N} \\
    \end{vmatrix}.
\end{equation}
The ratio of determinants in \cref{Seqn:GC_partition_detratio} can be simplified to yield an analytic expression,
\begin{equation}
    \mathcal{Z}_\mathrm{GCE}=(2\pi)^{N-1}\sum_{i=1}^N \tilde\alpha_i, \quad \tilde\alpha_i = \frac{e^{-\mu_i}}{\prod_{j\neq i}(\mu_j-\mu_i)}.
    \label{Seqn:GC_Z_analytic}
\end{equation}
Substituting \cref{Seqn:GC_Z_analytic} into the population constraint \cref{Seqn:GC_pop_constraint} yields
\begin{equation}
    \langle \Gamma_{nn} \rangle = -\frac{\partial}{\partial\mu_n}\ln\left(\sum_{i=1}^N\tilde\alpha_i\right) =\frac{ \tilde\alpha_n-\sum_{i \neq n}\frac{\tilde\alpha_i+\tilde\alpha_n}{\mu_i-\mu_n}}{\sum_i \tilde\alpha_i}. 
\end{equation}
Equating the populations to the Gibbs state gives a set of $N$ coupled equations for the Lagrange multipliers,
\begin{equation}
    \frac{ \tilde\alpha_n-\sum_{i \neq n}\frac{\tilde\alpha_i+\tilde\alpha_n}{\mu_i-\mu_n}}{\sum_i \tilde\alpha_i}=p_n, \quad n \in \{1,2,\cdots, N\},
    \label{eqn:mu_equations}
\end{equation}
which can be numerically solved for $\{\mu_n\}_{n=1}^N$. The solution is well approximated by
\begin{equation}
    \mu_i \approx -F(\beta)p_i + \frac{1}{p_i},
    \label{eqn:mu_ansatz}
\end{equation}
where
\begin{equation}
    \lim_{\beta\rightarrow 0}F(\beta) = N, \ \mathrm{and} \  \lim_{\beta\rightarrow \infty}F(\beta) = 1,
\end{equation}
which is supported by numerical comparisons in \cref{fig:mus}, where we have set $F(\beta)=\left(\sum_i p_i^2\right)^{-1}$, to match the limits given above.

\subsection{R\'enyi divergence constraint}
\label{SI:Renyi}

The R\'enyi divergence constraint can generate the Scrooge ensemble as a maximum entropy distribution. The constraint and corresponding observable are given by
\begin{gather}
    \langle D_\alpha(\Gamma\|\rho)\rangle_P = C(\rho), \\
    f(\Gamma) \rightarrow D_\alpha(\Gamma\|\rho)=\frac{1}{\alpha-1}\ln\mathrm{Tr}\{\Gamma\rho^{1-\alpha}\}, \quad \alpha \neq1,\label{Seqn:Renyi_constraint}
\end{gather}
where $C(\rho)$ is the constraint function, and $D_\alpha(\Gamma\|\rho)$ is the $\alpha$-R\'enyi divergence of wavefunction $\Gamma$ from the density operator $\rho$. Substituting \cref{Seqn:Renyi_constraint} into \cref{Seqn:P_constraints} yields the maximum entropy distribution,
\begin{equation}
    P_{\alpha\mu}(\Gamma)= \frac{(\mathrm{Tr}\{\Gamma\rho^{1-\alpha}\})^{-\mu}}{\int[d\Gamma](\mathrm{Tr}\{\Gamma\rho^{1-\alpha}\})^{-\mu}}, \quad \alpha \neq 1,
\end{equation}
where $\mu$ is the Lagrange multiplier enforcing the constraint. We retroactively determine the constraint function $C(\rho)$ by solving
\begin{align}
    C(\rho) &= \int[d\Gamma] P_{\alpha\mu}(\Gamma)\ln\mathrm{Tr}\{\Gamma\rho^{1-\alpha}\}, \\
            &= \frac{\int_{\Delta^{N-1}}dx \ \left(\sum_i x_i p_i^{1-\alpha}\right)^{-\mu}\ln\left(\sum_i x_i p_i^{1-\alpha}\right)}{\int_{\Delta^{N-1}}dx \ \left(\sum_i x_i p_i^{1-\alpha}\right)^{-\mu}},
\end{align}
where we have integrated out the wavefunction phases to reduce to a population integral over the probability simplex $\Delta^{N-1}$.

Define the auxiliary function,
\begin{align}
    I(\mu, \{\lambda_i\}) &= \int_{\Delta^{N-1}}dx \ \left(\sum_i x_i \lambda_i^{-1}\right)^{-\mu}, \\
           &= \frac{1}{\Gamma(\mu)}\int_0^\infty dt\ t^{\mu-1}\int_{\Delta^{N-1}}dx \ e^{-t\sum_i x_i \lambda_i^{-1}}, \\
           &= \frac{1}{\Gamma(\mu)}\sum_i\frac{\int_0^\infty dt\ t^{\mu-N}e^{-t\lambda_i^{-1}}}{\prod_{j\neq i}(\lambda_j^{-1}-\lambda_i^{-1})}, \\
           &= \frac{\Gamma(\mu-N+1)}{\Gamma(\mu)}\sum_i\frac{\lambda_i^{-(N-\mu-1)}}{\prod_{j\neq i}(\lambda_j^{-1}-\lambda_i^{-1})}, \\
           &= \frac{\Gamma(\mu-N+1)}{\Gamma(\mu)} \left(\prod_l \lambda_l\right) \sum_i\frac{\lambda_i^{\mu-1}}{\prod_{j\neq i}(\lambda_i-\lambda_j)}
\end{align}
which we have solved using the following results:
\begin{gather}
    \left(\sum_i x_i \lambda_i^{-1}\right)^{-\mu} = \frac{1}{\Gamma(\mu)}\int_0^\infty dt\ t^{\mu-1}e^{-t\sum_i x_i \lambda_i^{-1}}, \label{Seqn:res1}\\
    \int_{\Delta^{N-1}}dx \ e^{-t\sum_i x_i \lambda_i^{-1}} = \frac{1}{t^{N-1}}\sum_i\frac{e^{-t\lambda_i^{-1}}}{\prod_{j\neq i}(\lambda_j^{-1}-\lambda_i^{-1})}, \label{Seqn:res2}\\
    \Gamma(z) = \int_0^\infty dt \ t^{z-1}e^{-t}. \label{Seqn:res3}
\end{gather}
The partition function is then given by
\begin{align}
    \mathcal{Z}_{\alpha\mu} &= \int[d\Gamma](\mathrm{Tr}\{\Gamma\rho^{1-\alpha}\})^{-\mu}, \\
    &=(2\pi)^{N-1}I(\mu, \{p_i^{\alpha-1}\}), \\
    &=(2\pi)^{N-1}\frac{\Gamma(\mu-N+1)}{\Gamma(\mu)}(\det\rho)^{\alpha-1}\sum_i w_i(\alpha, \mu), 
\end{align}
where we have defined
\begin{equation}
    w_i(\alpha,\mu) = \frac{p_i^{(\alpha-1)(\mu-1)}}{\prod_{j \neq i} (p_i^{\alpha-1}-p_j^{\alpha-1})}.
\end{equation}

The constraint function can then be written as
\begin{align}
    C(\rho) &=-\frac{\partial}{\partial\mu}\ln \mathcal{Z}_{\alpha\mu}, \\
            &= -\frac{\sum_i w_i(\alpha,\mu)\ln p_i}{\sum_i w_i(\alpha, \mu)} + \frac{1}{\alpha-1}\sum_{k=0}^{N-2}\frac{1}{k+\mu-N+1},
\end{align}
where the Scrooge ensemble corresponds to $\alpha=2$ and $\mu=N+1$, leading to,
\begin{equation}
    C(\rho) = Q(\rho) + H_N -1
\end{equation}
where
\begin{equation}
    Q(\rho) = -\sum_{k} \frac{p_k^N\ln p_k}{\prod_{i\neq k}(p_k-p_i)}
\end{equation}
is the subentropy, and $H_N = \sum_{k=1}^Nk^{-1}$ is the $N$-th harmonic number.
\section{The R\'enyi divergence with $\alpha=2$ is the unique self-consistent constraint}
\label{SI:uniqueness}

\begin{figure}
    \centering
    \includegraphics[width=\columnwidth]{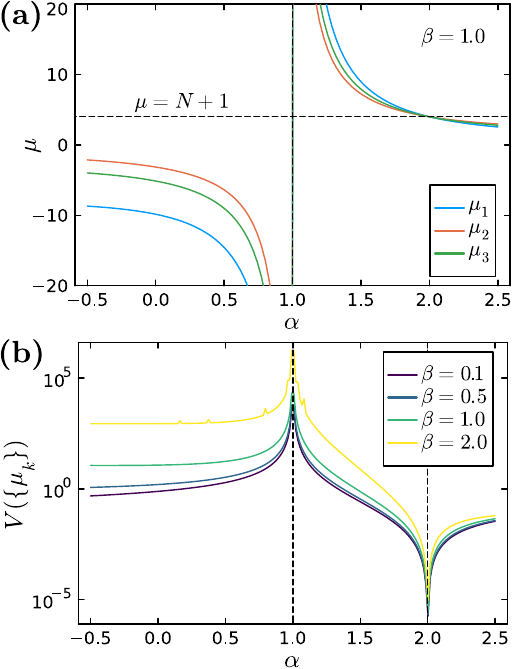}
    \caption{\textbf{There is a unique R\'enyi divergence  constraint $\langle D_\alpha(\Gamma\|\rho)\rangle_P=C(\rho)$ where the maximum entropy distribution has an average $\tilde{\rho}$ that is equal to $\rho$. (a)} The $N$ Lagrange multipliers $\{\mu_k\}$ obtained by numerically solving the $N$ population constraints $\tilde{\rho}=\rho$ (\cref{eqn:pop_alphamu}) as a function of $\alpha$. A valid solution requires that $\{\mu_k\}$ be identical, which is only satisfied at $\alpha=2$, where $\mu_k = N+1, \ \forall k$. \textbf{(b)} The variance of $\{\mu_k\}$ only vanishes at $\alpha=2$, for all density matrices and temperatures $\beta$. Results are calculated for a Hamiltonian spectrum $\sigma\{H\}=\{0,1,2\}$.}
    \label{fig:uniqueness}
\end{figure}

In this section, we prove that the $\alpha =2$ constraint is the unique R\'enyi divergence constraint of the form
\begin{equation}
    \langle D_\alpha(\Gamma\|\rho)\rangle_P = C(\rho),
\end{equation}
that is consistent with the Gibbs state (satisfying $\langle \Gamma \rangle_{P} = \rho, \ \forall\rho$). In other words, self-consistency requires that the density operator which we constrain the R\'enyi divergence on, to be equal to the density operator of the corresponding maximum entropy distribution. In the main text, we set $\alpha=2$ and $\mu=N+1$ to coincide with the Scrooge ensemble, without any a priori principle to select this constraint; however, the following uniqueness property proves that there is no other self-consistent alternative. The population of the $k$-th state in the distribution $P_{\alpha\mu}(\Gamma)$ is given by (proof in section \ref{SI:Renyi_operator} of the SI), 
\begin{align}
    \tilde{p}_{k}(\alpha,\mu)&:=\langle \Gamma_{kk} \rangle_{P_{\alpha\mu}} \\ 
    &= \frac{\lambda_{k\alpha} }{\mu-N}\frac{\partial}{\partial \lambda_{k\alpha} }\ln\sum_i\frac{\lambda_{i\alpha}^{\mu-1}}{\prod_{j \neq i} (\lambda_{i\alpha} -\lambda_{j\alpha} )} 
    \label{eqn:pop_alphamu}
\end{align}
where $\lambda_{k\alpha} = p_k^{\alpha-1}$, and we assume that $\alpha\neq1$. A parameter set $(\alpha,\mu)$ is consistent with the Gibbs state if the density operator corresponding to $P_{\alpha\mu}$ is equal to the density operator $\rho$ in the constraint (i.e., $\tilde{p}_{k}(\alpha,\mu)= p_k, \ \forall k, \rho$). Assuming that $(\alpha,\mu)$ is consistent with the Gibbs state, we can integrate the $N$ partial derivatives \cref{eqn:pop_alphamu} to yield,
\begin{equation}
    (\alpha-1)(\mu-N) + c = \ln\sum_i\frac{\lambda_{i\alpha}^{\mu-1}}{\prod_{j \neq i} (\lambda_{i\alpha} -\lambda_{j\alpha} )},
    \label{eqn:uniqueness}
\end{equation}
where $c$ is a constant that is independent of $\rho$. Since the left-hand side of \cref{eqn:uniqueness} is independent of $\rho$, the right-hand side must also be independent. Using identities from Lagrange interpolants, we have that 
\begin{equation}
    \sum_i\frac{\lambda_{i\alpha}^{\mu-1}}{\prod_{j \neq i} (\lambda_{i\alpha} -\lambda_{j\alpha} )} = 
    \begin{cases}
            0   \quad \mathrm{if} \quad  \mu = 0,1,2,\dots,N-1\\
            1 \quad \mathrm{if} \quad \mu = N\\
            \sum_i \lambda_{i\alpha} \quad \mathrm{if} \quad \mu = N+1\\
            \mathrm{nontrivial \ otherwise}
    \end{cases}
\end{equation}
The case $\mu = N+1$ is independent of $\rho$ only when $\alpha=2$, such that $\sum_i \lambda_{i\alpha} = \sum_i p_i = 1$. The cases $\mu = \{0,1,\dots,N\}$ correspond to spurious solutions from the differentiation and integration of \cref{eqn:pop_alphamu}, as can be confirmed by numerical solutions. \Cref{fig:uniqueness} shows that the variance of the $N$ Lagrange multipliers obtained by solving the populations $\tilde{p}_k(\alpha,\mu_k)=p_k$ only vanishes at $\alpha=2$, where $\mu_k=N+1, \ \forall k$.

Finally, we consider the remaining $\alpha=1$ case. This case produces no constraint on the wavefunction distribution because the average divergence is invariant under $P$ ($\langle D_1(\Gamma\|\rho)\rangle_P = S_\mathrm{vN}(\rho), \ \forall P$).

Therefore, the unique solution that is consistent with the Gibbs state corresponds to the constraint $\langle D_2(\Gamma\|\rho)\rangle_P = \langle H_A(\rho) \rangle_A$. Consequently, we have proven that any constrained R\'enyi divergence between the density operator and its composing wavefunctions must yield the Scrooge ensemble in a maximum entropy procedure.

\section{Density operators of the maximum entropy distributions}
\label{SI:density_op}

In this section, we derive the density operators of the maximum entropy distributions under the energy, Gibbs-state, and R\'enyi divergence constraints.

\subsection{Energy constraint}
\label{SI:ECE_operator}

The populations of the ECE density operator can be expressed as derivatives of the ECE partition function,
\begin{align}
    p_{k}^\mathrm{ECE}&:=\langle \Gamma_{kk} \rangle_{P_\mathrm{ECE}}, \\ 
    &= \int [d\Gamma]\Gamma_{kk}P_\mathrm{ECE}(\Gamma), \\
    &= \frac{\int [d\Gamma] \Gamma_{kk} \exp(-\beta'\bar{H}(\Gamma))}{\int [d\Gamma] \exp(-\beta'\bar{H}(\Gamma))}, \\
    &= -\frac{\partial}{\partial (\beta' E_k)}\ln\left(\int [d\Gamma] \exp(-\beta'\mathrm{Tr}\{H\Gamma\})\right).
\end{align}
Substituting in the analytic form of the partition function \cref{Seqn:Z_analytic} yields
\begin{align}
    p_{k}^\mathrm{ECE} &= -\frac{\partial}{\partial (\beta' E_k)}\ln\left(\sum_{i=1}^N \alpha_i\right),  \\
    &= \frac{\alpha_k + \sum_{i\neq k} \frac{\alpha_k  + \alpha_i}{\beta'E_k - \beta'E_i}}{\sum_{i=1}^N \alpha_i},
\end{align}
where
\begin{equation}
    \alpha_i = \frac{e^{-\beta'E_i}}{\prod_{j\neq i}(\beta'E_j-\beta'E_i)},
\end{equation}
as presented in the main text.

\subsection{Gibbs-state constraint}
By construction, the GCE density operator is the Gibbs state.

\subsection{R\'enyi divergence constraint}
\label{SI:Renyi_operator}

The populations of the maximum entropy distribution under the R\'enyi divergence constraint are given by 
\begin{align}
    &\tilde{p}_{k}(\alpha,\mu):=\langle \Gamma_{kk} \rangle_{P_{\alpha\mu}}, \\ 
    &= \int [d\Gamma] \Gamma_{kk} P_{\alpha\mu}(\Gamma), \\
    &= \frac{\int [d\Gamma]\Gamma_{kk} \left(\mathrm{Tr}\{\Gamma\rho^{1-\alpha}\}\right)^{-\mu}}{\int [d\Gamma]\left(\mathrm{Tr}\{\Gamma\rho^{1-\alpha}\}\right)^{-\mu}}, \\
    &= \frac{(2\pi)^{N-1}}{Z_{\alpha\mu}}\int_{\Delta^{N-1}}dx \ x_k \left(\sum_i x_i\lambda_{i\alpha}^{-1}\right)^{-\mu}, \quad \lambda_{i\alpha} = p_i^{\alpha-1} \\
    &= \frac{(2\pi)^{N-1}}{Z_{\alpha\mu}\Gamma(\mu)}\int_0^\infty dt\ t^{\mu-1} \int_{\Delta^{N-1}}dx \ x_k \ e^{-t\sum_i x_i \lambda_{i\alpha}^{-1}}, \\
    &= \frac{\displaystyle \left(\mu-1 + \sum_{i\neq k}\frac{\lambda_{k\alpha}}{\lambda_{i\alpha}-\lambda_{k\alpha}}\right)w_k+\sum_{i\neq k} w_i\frac{\lambda_{k\alpha}}{\lambda_{i\alpha}-\lambda_{k\alpha}}}{\displaystyle (\mu-N)\sum_i w_i}, \\
    &= \frac{\lambda_{k\alpha} }{\mu-N}\frac{\partial}{\partial \lambda_{k\alpha} }\ln\sum_iw_i(\alpha,\mu),
\end{align}
where we have used \cref{Seqn:res1}, \cref{Seqn:res2}, \cref{Seqn:res3}, and defined
\begin{equation}
    w_i(\alpha,\mu) =\frac{\lambda_{i\alpha}^{\mu-1}}{\prod_{j \neq i} (\lambda_{i\alpha} -\lambda_{j\alpha} )}= \frac{p_i^{(\alpha-1)(\mu-1)}}{\prod_{j \neq i} (p_i^{\alpha-1}-p_j^{\alpha-1})}.
\end{equation}

\section{Ensemble entropy of ECE, GCE, and Scrooge ensembles}
\label{SI:entropy}

\begin{figure}
    \centering
    \includegraphics[width=\columnwidth]{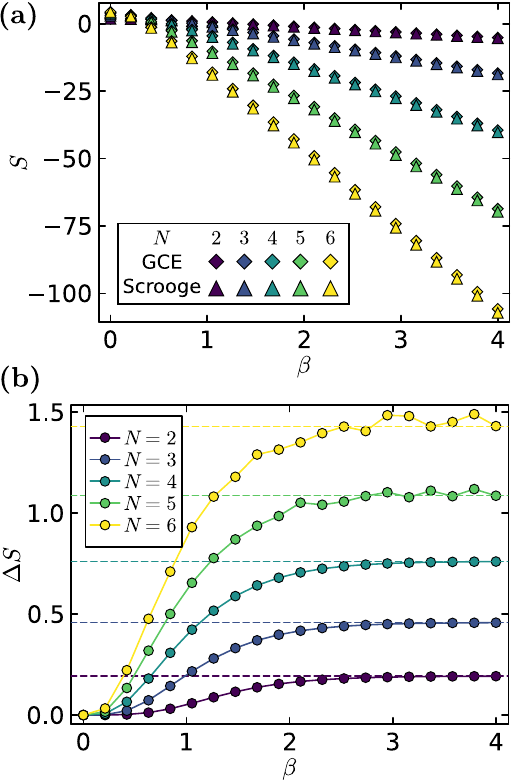}
    \caption{\textbf{The Scrooge ensemble only converges to the GCE in the infinite temperature limit. (a)} The ensemble entropy of the GCE ($S_\mathrm{GCE}$) is always greater than or equal to the entropy of the Scrooge ensemble ($S_\mathrm{Scr}$). \textbf{(b)} The entropy difference $\Delta S = S_\mathrm{GCE}-S_\mathrm{Scr}$ approaches the limit \cref{eqn:entropy_diff_limit} (dashed lines) at low temperatures. The difference $\Delta S$ increases with $N$, showing that convergence is not achieved in the thermodynamic limit. Results are calculated for Hamiltonian spectra $\sigma\{H\} = \{k\}_{k=0}^{N-1}$.}
    \label{fig:entropy}
\end{figure}

Here, we present and compare the ensemble entropies of the candidate thermal distributions. The ensemble entropy is defined as
\begin{equation}
    S_P(P) = -\int[d\Gamma]P(\Gamma)\ln P(\Gamma).
\end{equation}
The entropies of the ECE, GCE, and Scrooge ensemble are respectively given by
\begin{gather}
    S_\mathrm{ECE} = \beta'U + \ln \mathcal{Z}_\mathrm{ECE}, \\
    S_\mathrm{GCE} = \mathrm{Tr}\{M\rho\} + \ln \mathcal{Z}_\mathrm{GCE}, \\
    S_\mathrm{Scr} = (N+1)(Q(\rho) + H_N -1) + \ln \mathcal{Z}_\mathrm{Scr},
\end{gather}
where 
\begin{align}
    \mathcal{Z}_\mathrm{Scr} &= \int[d\Gamma] (\mathrm{Tr}\{\Gamma\rho^{-1}\})^{-(N+1)}, \\
                             &= \frac{(2\pi)^{N-1}\det\rho}{N!}.
\end{align}

\subsection{The Scrooge ensemble does not maximize entropy under a Gibbs-state constraint}

Here, we show that the Scrooge ensemble does not converge to the GCE at low temperatures or in the thermodynamic limit. The entropy difference at low temperatures is given by
\begin{multline}
    \lim_{\beta\rightarrow\infty}(S_\mathrm{GCE}-S_\mathrm{Scr})  \\ =\lim_{\beta\rightarrow\infty}\Big[\ln \left(\frac{\mathcal{Z}_\mathrm{GCE}}{\mathcal{Z}_\mathrm{Scr}} \right)+ \mathrm{Tr}\{M\rho\} \\ - (N+1)(Q(\rho) + H_N -1)\Big] 
\end{multline}
\begin{multline}
    \hspace{1.5cm}= \ln \left(\lim_{\beta\rightarrow\infty}\frac{\mathcal{Z}_\mathrm{GCE}}{\mathcal{Z}_\mathrm{Scr}} \right) + (N-1) \\ - (N+1)(H_N -1)
\end{multline}
where we have used that $\lim_{\beta\rightarrow\infty}Q(\rho)=0$, and $\lim_{\beta\rightarrow0}\mathrm{Tr}\{M\rho\}=\sum_i\mu_ip_i=N-1$.

To evaluate the limit of the partition functions, we use \cref{eqn:mu_ansatz} with
\begin{equation}
    p_1\approx1, \ p_i\approx 0,  \ i> 1 \implies \mu_1 \approx 0, \ \mu_i \approx \frac{1}{p_i} \gg 1,
\end{equation}
which allows us to truncate the sum in $\mathcal{Z}_\mathrm{GCE}$ and evaluate the limit as
\begin{align}
    \lim_{\beta\rightarrow\infty}\frac{\mathcal{Z}_\mathrm{GCE}}{\mathcal{Z}_\mathrm{Scr}} &= \frac{N!}{\det\rho}\sum_{i=1}^N  \frac{e^{-\mu_i}}{\prod_{j\neq i}(\mu_j-\mu_i)}, \\
    &\approx \frac{N!}{\det\rho} \frac{e^{-\mu_1}}{\prod_{j\neq 1}(\mu_j-\mu_1)}, \\
    &\approx N!.
\end{align}
Therefore, the difference in entropy between the Scrooge ensemble and GCE at low temperatures is given by
\begin{equation}
    \lim_{\beta\rightarrow\infty}(S_\mathrm{GCE}-S_\mathrm{Scr}) = \ln N!+2N - (N+1)H_N >0,
    \label{eqn:entropy_diff_limit}
\end{equation}
proving that the entropies do not converge at low temperature. Furthermore, the entropy difference grows linearly with $N$, proving that convergence is not achieved in the thermodynamic limit,
\begin{equation}
    \lim_{N\rightarrow\infty}\lim_{\beta\rightarrow\infty}(S_\mathrm{GCE}-S_\mathrm{Scr}) = N(1-\gamma),
\end{equation}
where $\gamma$ is the Euler-Mascheroni constant. \Cref{fig:entropy} shows that the numerically calculated entropy difference indeed converges to our derived limit.

\subsection{Entropy of $\rho$-distortion from Haar to Scrooge ensemble}
We remark that the entropy difference between the Haar ensemble and Scrooge ensemble can be written in terms of the average R\'enyi divergence,
\begin{equation}
    S_\mathrm{Haar} - S_\mathrm{Scr} = \ln\left(\frac{N}{\det\rho}\right) - (N+1)\langle D_2(\Gamma\|\rho)\rangle_P,
    \label{Seqn:DeltaS}
\end{equation}
where $S_\mathrm{Haar} = \ln \mathcal{Z}_\mathrm{Haar}$, and $\mathcal{Z}_\mathrm{Haar} = \int[d\Gamma] = \frac{(2\pi)^{N-1}}{(N-1)!}$. \Cref{Seqn:DeltaS} is  the entropy difference incurred by a $\rho$-distortion i.e., the method used by Jozsa et al. to derive the Scrooge ensemble from deforming the Haar ensemble.\cite{Jozsa1994}

\section{Ground state condensation of the ECE in the thermodynamic limit}
\label{SI:condensation}

\begin{figure}
    \centering
    \includegraphics[width=\columnwidth]{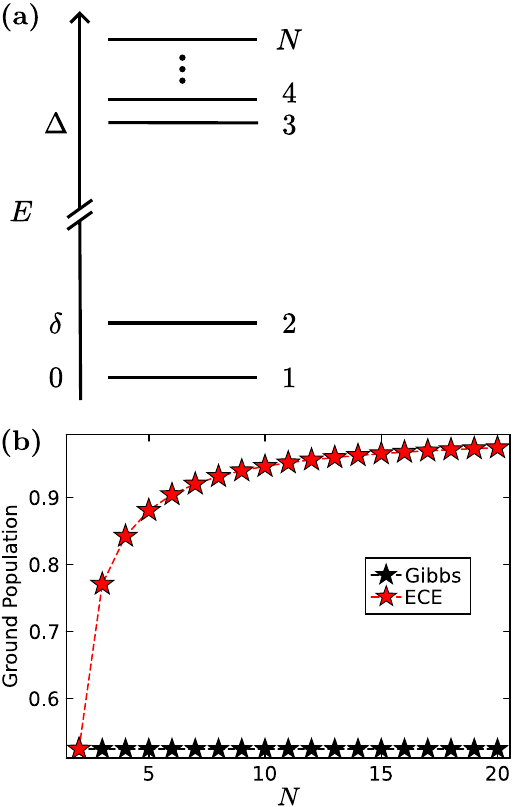}
    \caption{\textbf{The ECE exhibits ground state condensation in the thermodynamic limit.} \textbf{(a)} The energy level diagram of an $N$-level system with two low-energy states at $0$ and $\delta$, and a set of high-energy levels above $\Delta \gg \delta$. \textbf{(b)} The ground state population in the ECE (red stars) increases with $N$, even though it should be unaffected by the inclusion of thermally-inaccessible levels, like the Gibbs state (black). Results are calculated for $\delta=0.1$, $\Delta = 100$, and $\beta = 1$.}
    \label{fig:truncation}
\end{figure}

The ECE does not agree with the Gibbs state because it exhibits a condensation of the ensemble into the ground state in the thermodynamic limit. Ground state condensation has been previously reported in the quantum microcanonical ensemble, defined as the ensemble of all wavefunctions with a fixed energy expectation value.\cite{Fine2009, Hantschel2011} Here, we show that condensation also occurs in the ECE by considering the case of adding thermally inaccessible energy levels to a given system (\cref{fig:truncation}a). A physical model should be independent of thermally inaccessible levels; however, \cref{fig:truncation}b shows that the ground state population increases with each additional high-energy level, eventually producing a thermal state completely in the ground state. The condensation occurs because the low-energy subspace becomes saturated by wavefunctions with near-unity ground state coefficients.

\end{document}